# Development of a low-cost monitor for radon detection in air

Soraia Elísio[1,2] and Luis Peralta[1,2]

[1] Faculdade de Ciências da Universidade de Lisboa

[2] Laboratório de Instrumentação e Física Experimental de Partículas

**Abstract:** An active device for radon detection in the air was developed. The monitor operates in pulse counting mode for real-time continuous measurements. The presented prototype has a relatively simple design made of low-price and easy to acquire components which made it possible to develop an inexpensive device. The device used as a sensor, the SLCD-61N5 Si-PIN planar photodiode, which has an area of 9.67×9.67 mm$^2$, is sensitive to alpha particles. An Arduino Uno microcontroller was used as a data acquisition system. Signals were observed when placing an $^{241}$Am or $^{226}$Ra source near the sensor. The sensor's sensitivity has small bias dependency and the device can operate even at modest voltage. As a result of a one-month test in a radon-rich atmosphere, a positive high correlation (Pearson's r equal to 0.977) was obtained between our prototype and a Geiger-Müller detector.

## 1. Introduction

Exposure to ionizing radiation from natural sources is continuous and inevitable for all living organisms. According to UNSCEAR [1], radon once accumulated in homes presents the main contribution to exposure from natural sources and consequently is a risk factor for health. However, the main concern is due to the inhalation and deposition of the solid radon progeny in the respiratory tract. Since alpha particles interact strongly with matter, they deposit all their energy in a very small volume of bronchial tissue. The cumulative dose of alpha radiation to cells can lead to the development of lung cancer [2-5].

There are many different instruments and techniques available for radon detection and quantification in the air [6]. The methodology essentially involves the interaction of either charged particles or gamma-rays with the sensor material. The most common means for the measurement of radon is through the detection of alpha emission from the decay of $^{222}$Rn (5.49 MeV), $^{218}$Po (6.00 MeV) and $^{214}$Po (7.69 MeV). Among the wide range of available devices, the passive detectors have been used in large-scale radon surveys, such as the CR-39 detectors [7], due to their simplicity, lightweight and low price. However, active detectors working with electrostatic collection of the charged radon progeny have been developed by many authors for radon measurements [8-12].

In this paper, we present an active detector for radon in the air using a low-cost silicon PIN photodiode (cost of approximately ten euros) as the main sensor. In addition, we also use a flexible Arduino Uno microcontroller as the data acquisition system. The device's operation is based on the counting of alpha particle's interactions with the photodiode sensitive volume, due primarily to the decay of $^{218}$Po and $^{214}$Po. No energy analysis is made. This work presents the results of experiments carried out in a laboratory environment with radon and its progeny exhaled from rocks. Measurements were performed simultaneously for the purpose of comparison with a Geiger-Müller detector. Our objective was to develop a radon detection device that can be used by university students in field trips or academic work.

## 2. The SLCD-61N5 photodiode

The Silonex SLCD-61N5 (now Advanced Photonix) [13] is a low-cost planar silicon PIN photodiode (cost of approximately 10 euros). The active area of the photodiode is 9.67×9.67 mm$^2$ and the reverse breakdown voltage is -20 V. The main application of this photodiode is the detection of visible and infrared radiation, but the very thin window makes it a candidate for the detection of alpha and beta particles. Unfortunately, the window thickness and depletion layer thickness are not provided by the manufacturer. Gugliermetti et al [14] estimated the window to be of the order of 100 nm. The sensor's total thickness is 0.4 mm, including the back support.

**2.1 Detecting alpha particles with the SLCD-61N5 photodiode**

To assess the sensor's capability to detect alpha radiation, tests were made using an $^{214}$Am and a $^{226}$Ra point-like alpha sources. For the $^{241}$Am source the most intense alpha peaks are the 5.49 MeV (84.5%) and 5.44 MeV (13.0%). For the $^{226}$Ra source the most intense alpha-peaks are the 4.78 MeV (94.4%) from the $^{226}$Ra decay, 5.49 MeV (99.9%) from the $^{222}$Rn decay, 6.00 MeV (100%) from $^{218}$Po, 7.69 MeV (100%) from $^{214}$Po and 5.30 MeV (100%) from $^{210}$Po [15]. For charge collection, the sensor was connected to a Hamamatsu H4083 charge amplifier. The signal was then fed to an Ortec 575A amplifier. The amplifier output was connected to an Amptek 8000A Multichannel Analyser. For response comparison purposes, the Hamamatsu S3590-09 photodiode [16] was used. The S3590-09 is a windowless photodiode with 10×10 mm$^2$ active area, for particle detection and has been used in works for alpha-particle detection from radon decay [17]. The electronic chain connected to the S3590-09 was the same used for the SLCD-61N5. Both photodiodes were biased with -18 V using a Tenma 72-10480 DC power supply. The detectors were placed inside a light-tight aluminium box. For data collection, the radioactive sources were placed directly on top the surface of the detector, being the distance between the photodiode's surface and the source's surface being approximately 1 mm. The air inside the box was not evacuated. The spectra acquired over a period of 10 minutes are presented in figure 1.

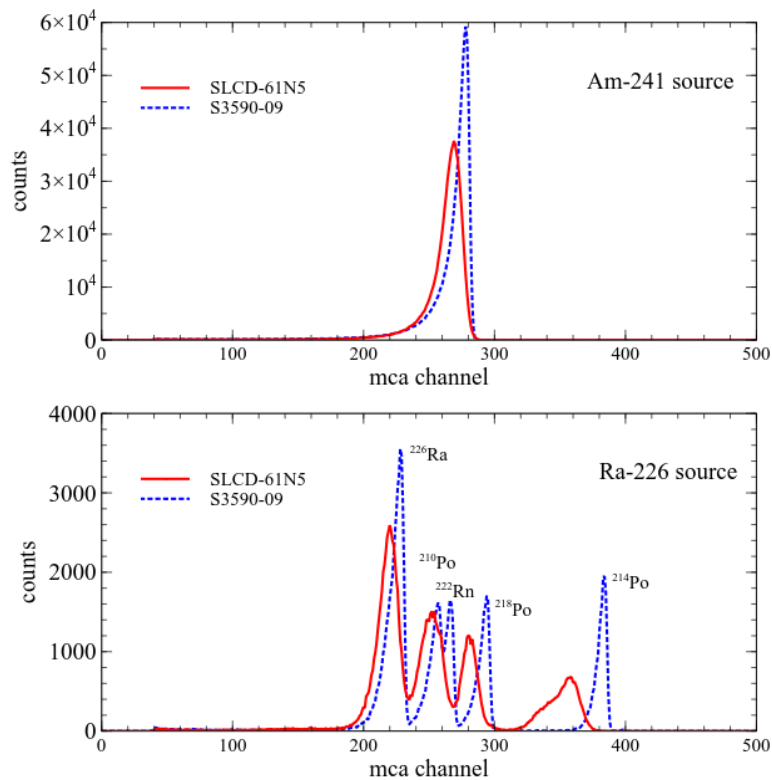

Figure 1. Alpha particle spectra of two radioactive sources ($^{241}$Am and $^{226}$Ra) acquired with the Silonex SLCD-61N5 and the Hamamatsu S3590-09 photodiodes polarized at -18 V.

The spectra obtained with the SLCD-61N5 photodiode clearly follows the same pattern as the spectra acquired with the S3590-09 photodiode. All expected alpha peaks in the $^{226}$Ra source are present in the spectrum acquired with the detector S3590-09, while the SLCD-61N5 detector does not resolve the 5.30 MeV peak ($^{210}$Po) from 5.49 MeV peak ($^{222}$Rn). In fact, due to the entrance window, the energy resolution of the SLCD-61N5 photodiode is worse than the one from the S3590-09 photodiode. In the $^{226}$Ra source spectrum the lower counting rate of the alpha-peaks from the decay of $^{222}$Rn, $^{218}$Po, $^{214}$Po and $^{210}$Po can be attributed to the escape of the radon gas from the thin source deposit sample, decreasing the amount of that element present in the source.

To test the SLCD61N5 count rate stability, acquisitions were made at different biases (from 0 to -18 V) using the $^{241}$Am source. The count rate variability was measured to be less than 1%. The signal amplitude increased by 40% from 0 to -9 V, and 2-3% in the range -9 to -18 V.

## 2.2 Radon detection with the SLCD61N5 photodiode

The next step was to test the SLCD61N5 photodiode ability to detect alpha particles emitted from radon and radon progeny ($^{218}$Po and $^{214}$Po). The photodiode was placed inside a 5.5×9.0×11.5 cm$^3$ aluminium box (HAMMOND 1550C), with a rubber gasket on the lid. The rubber gasket minimizes the radon leakage into the lid junction. The box was connected through a hose to a container holding uranium oxide-rich rocks used as radon source and the gas diffuses through the hose into the test box. To have a concentration equilibrium between the container and the box holding the photodiode, tests were conducted only a day after connecting the two boxes. Alpha spectra were then acquired for one hour each, at several photodiode biases. The spectra acquired at -5, -9 and -18 V are presented in figure 2. After the $^{222}$Rn decay the resulting $^{218}$Po ion can adhere to aerosol particles present in the box [6,8-12,17]. Some $^{218}$Po can still reach the detector and adhere to the surface. Alpha particles resulting from the decay of these $^{218}$Po (or descendent $^{214}$Po) may penetrate directly into the detector and give rise to peaks clearly seen in the spectra of figure 2. Alpha particles arriving to the detector after crossing a thickness of air give rise to the continuous part of the spectrum. A low threshold was set to eliminate signals due to beta particles from $^{214}$Pb and $^{214}$Bi decays. The total number of counts for the acquired spectra at -9 and -18 V remain constant within 2%. The effect of the more negative bias was the slight increase in the signal amplitude, compatible with the previous measurements made with the solid $^{241}$Am source.

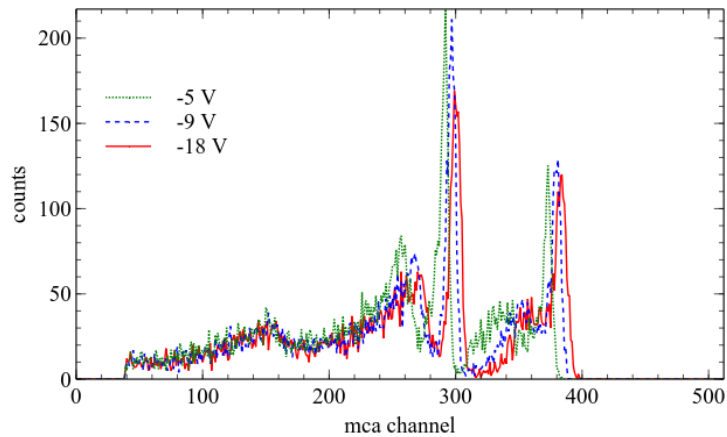

Figure 2. Radon spectra obtained with the SLCD-61N5 photodiode at -5, -9 and -18 V bias.

## 3. Prototype development

The developed radon monitor is divided into four main parts: the sensitive sensor (SLCD-61N5) and preamplifier, the amplification chain, the signal discriminator, the signal formatting and the data acquisition system. Figure 3 shows a block diagram of the system design. The anode of the photodiode is negatively biased with respect to the ground with -18 V. This value can be obtained from 2 standard 9 V batteries, enabling an easy stand-alone operation of the device. When a charged particle hits the sensor, a small current is produced with intensity proportional to the deposited energy. The current signal produced by the photodiode is converted to voltage by a transimpedance amplifier using a high impedance, low power consumption LF442 operational amplifier. Signal amplification is made by two LF442 operational amplifiers in voltage inverting configuration providing a gain of 2500 and discriminated by a LM311 comparator. The analogue signals are then converted into logic pulses with fixed amplitude, when the signal at the amplifier output is above a threshold. The threshold was experimentally adjusted so that the signal from a $^{241}$Am source, used as reference, was accepted. Before entering the Arduino microcontroller, the signal is formatted to a square signal of 4 V amplitude and 150 µs width by a NE555 timer, working in inverse monostable configuration.

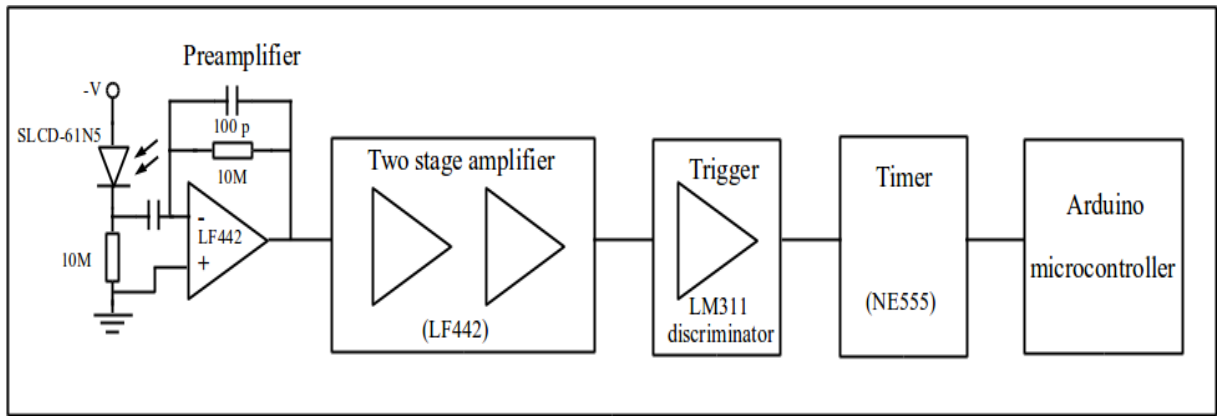

Figure 3. Block diagram of the detection system.

The microcontroller scores and stores the number of counts in a defined time interval. The scoring of the asynchronous hits is made via an interrupt to the microcontroller. The number of hits in each time lapse is written into a file in the computer along with the timestamp.

A picture of the assembled radon monitor prototype is depicted in figure 4. An aluminium box fabricated by HAMMOND, with 22.2×14.5×5.5 cm$^3$ is used to contain the full setup. Holes were drilled on the box side to allow the gas to enter by diffusion. The box interior was painted matt black. A labyrinth formed by two black slices were placed in the middle of the box preventing exterior light from reaching the sensor.

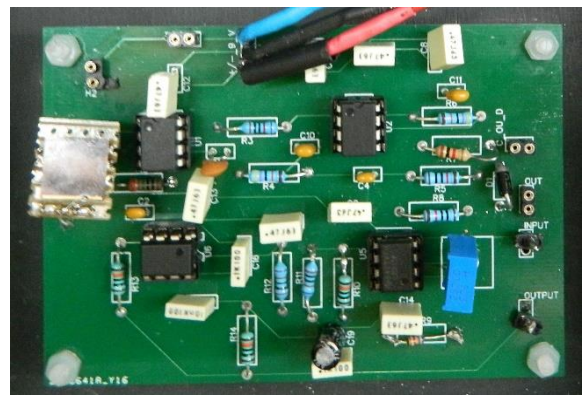

Figure 4. The prototype board. On the left sits the Silonex SLCD-61N5 sensor.

## 3.1 Tests with $^{241}$Am source

To test the system for alpha particles detection an $^{241}$Am alpha source was used. This source emits alpha-particles with energy of 5.48 MeV smaller than the 6.00 and 7.69 MeV alpha-particles emitted by the $^{218}$Po and $^{214}$Po radon progeny. As mentioned before, the detection of $^{241}$Am alpha-particles will be useful to set the low threshold for alpha-particle detection and noise rejection. After setting the threshold the number of counts without source was zero.

Figure 5a shows a typical amplifier output signal in absence of incident charged particles, while figure 5b shows a typical signal when a $^{241}$Am source is placed on the top of the sensor. Pulses with amplitude in the range 0.40 V to 1.8 V were observed. The trigger threshold was thus set below 0.40 V. The timer pulse width is set to 150 µs (figure 5c), enough not to be missed by the Arduino.

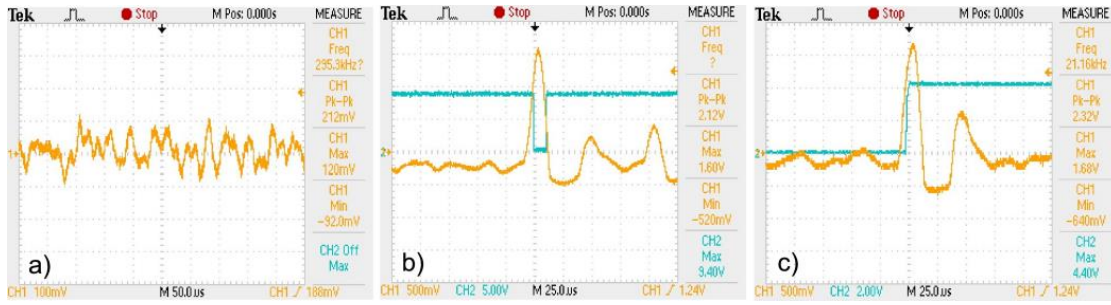

Figure 5. Oscilloscope signals at a) amplifier output when no alpha particles hit the sensor; b) amplifier and trigger output when alpha-particles from an [241]Am source hit the sensor; c) amplifier and timer output when alpha-particles from an [241]Am source hit the sensor;

## 4. Prototype radon tests

To test the system for radon detection under laboratory conditions the prototype was placed inside an acrylic chamber 20×21×30 cm$^3$. A GM-10 Geiger-Müller detector (from Black Cat Systems [18]) was placed inside the chamber along with the prototype. The chamber was connected through a hose to the container with uranium oxide rich rocks used as radon source. Temperature, atmospheric pressure and relative humidity were monitored using a BME280 sensor [19] connected to an Arduino.

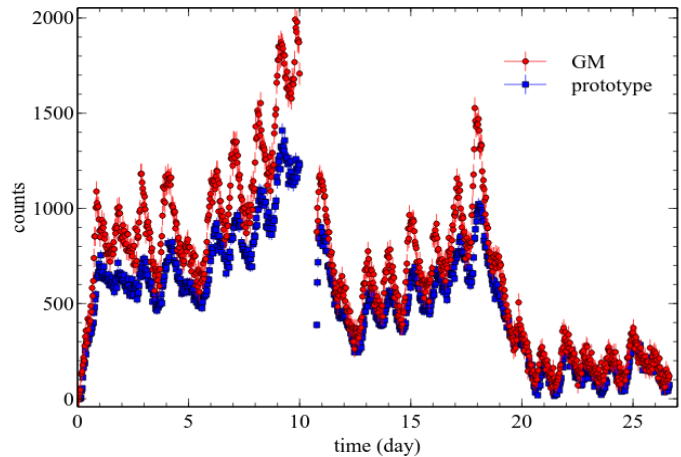

Figure 6. Long-term data collection for the radon monitor prototype and GM control. The prototype sensor is biased with -18 V. Each data point corresponds to one-hour acquisition. The GM data have been background subtracted.

An acquisition was performed with the prototype and GM detector inside the test box filled with radon. Figure 6 presents the counts in one-hour intervals for both the radon prototype and the Geiger-Müller detector. The GM counts are subtracted by the background average value in one hour. An acquisition interruption, due to a general power failure, took place for a few hours on the 10[th] acquisition day. As it can be seen from figure 6 the number of counts given by the prototype follows a similar pattern to that of the Geiger-Müller detector over the measurement period. On the graph regular counting oscillations are observed. These oscillations are related to the day/night temperature variations seen in figure 7.

To further investigate the Prototype-Geiger count correlation a scatter plot was made (figure 8). A Pearson correlation test was then applied to 931 points and a Pearson's r value equal to 0.977 obtained. This R value corresponds to p<0.001 indicating a high positive correlation between the data of our detector and those of the GM-10 counter.

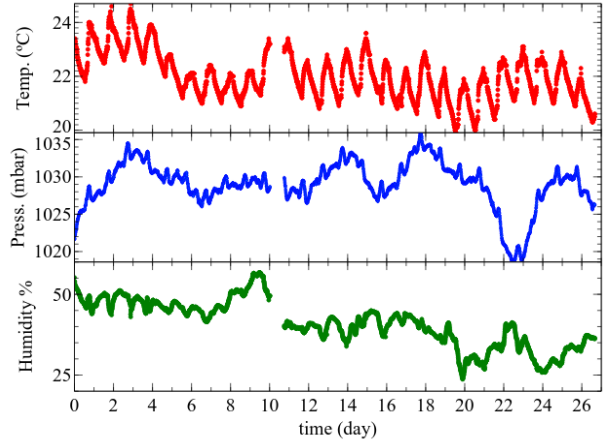

Figure 7. Temperature, barometric pressure and relative humidity measured with a BME280 sensor connected to an Arduino UNO microcontroller. The sensor has an accuracy of ±1 ºC for temperature measurement, ±1 mbar for barometric pressure measurement and ±3% for relative humidity measurement [19].

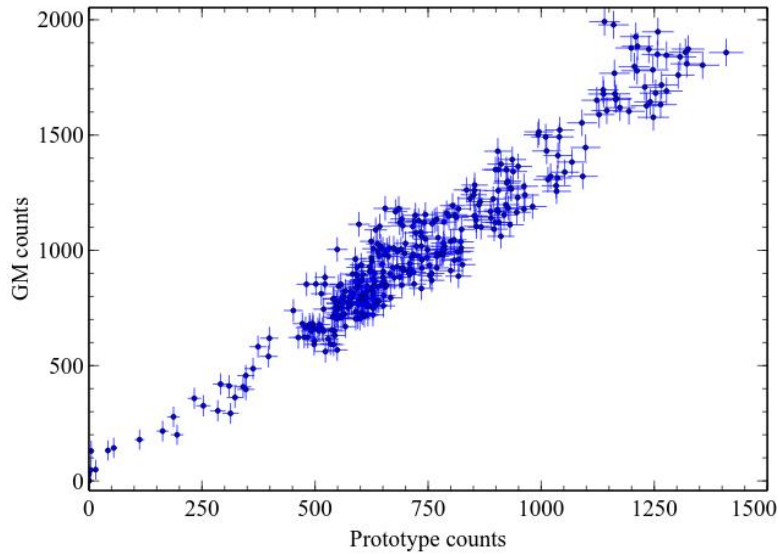

Figure 8. GM counts vs Prototype counts. A clear linear correlation between both counts is observed (Pearson's r equal to 0.977).

## 5. Radon concentration assessment

To roughly assess the radon concentration inside the test chamber, a Canberra 3×3 inch$^2$ NaI(Tl) detector, with a preamplifier Canberra Model 2007 was placed in contact with one side of the chamber (figure 9). The signal from the preamplifier was fed to an Ortec 575A amplifier and the output signal acquired by an Amptek 8000A Multichannel Analyser. A lead brick wall, 10 cm thick, was put between the test chamber and the rock's container, to greatly decrease the direct gamma radiation coming from the container. The energy calibration of the detector was made using $^{137}$Cs, $^{60}$Co and $^{22}$Na radioactive sources. A one-day acquisition of background radiation was made. The radon was then allowed to diffuse from the container to the test chamber for two days for radon concentration to build-up in the chamber. An acquisition of the gamma radiation spectrum from the chamber was then made for 6 hours.

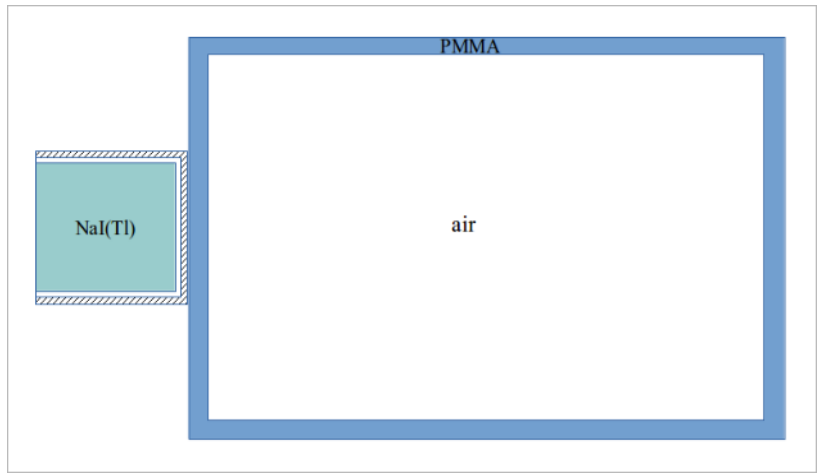

Figure 9. Simulated setup in Penelope Monte Carlo to compute the NaI(Tl) detector's efficiency for the $^{214}$Bi 609 keV gamma-peak. The chamber has 10 mm thick PMMA walls and is filled with air at standard pressure. The NaI(Tl) crystal is wrapped in a MgO reflector covered by an aluminium case. The MgO reflector and case thickness dimensions are not to scale in the figure.

In figure 10 the count rate spectra of background radiation and radiation from radon chamber are presented. The signal spectrum obtained by subtracting the radon chamber count rate from the background radiation is also presented in figure 10. A clearly demonstrated increase of the count rate due to the gamma peaks from the $^{214}$Pb and $^{214}$Bi, both being nuclides $^{222}$Rn descendants. In the signal spectrum of figure 10, the 609 keV gamma-peak from the $^{214}$Bi decay is well isolated and can be used to assess the nuclide activity inside the test chamber, where, it is possible to consider that secular activity equilibrium is achieved between $^{222}$Rn and its descendants. Thus, the determination of the $^{214}$Bi activity using the 609 keV gamma-peak will enable the determination of the radon activity in the chamber. The $^{214}$Bi activity can be obtained using the equation [20]

$$A = \frac{n_{214Bi}}{\epsilon} Br$$

where $n_{214Bi}$ is the 609 keV gamma-peak count rate, Br the peak branching ratio (0.461 for this peak [15]) and ε the efficiency for peak detection. The $n_{214Bi}$ count rate was obtained by integrating the count rate in a region of interest of ±3σ relative to the 609 keV peak-centroid. The detection efficiency for the 609 keV gamma-peak was obtained from the Penelope Monte Carlo simulation [21-23] of the setup depicted in figure 9. The detector was modelled as a 3×3 inch$^2$ NaI crystal, wrapped in a MgO 0.185 cm thick reflector inside a 0.05 cm thick aluminium case [24]. The chamber walls are 10 mm thick PMMA and the chamber is filled with air at standard atmospheric pressure. The 609 keV source points were uniformly distributed inside the chamber. The emission distribution of gamma photons was made isotropic. An overall efficiency of (7.2±0.2)×10$^{-3}$ for the detection of total absorption peak was obtained from the simulation of 10$^8$ events. Using this efficiency, the radon concentration values inside the chamber were evaluated to be in the 10 to 100 kBq/m$^3$ range, for prototype counts in the range of 100 to 1000 counts per hour.

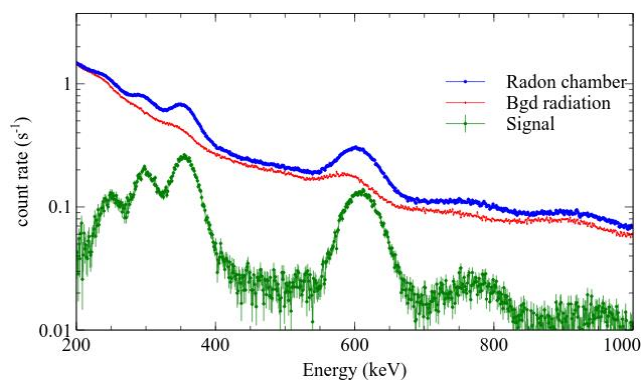

Figure 10. Radon chamber and background count rate spectra. The radon chamber spectrum was acquired over a period of 6 hours. The signal spectrum is obtained from the subtraction of radon chamber and background spectra.

## 6. Conclusion

The SLCD-61N5 is a low-cost planar photodiode with an active area of 9.67×9.67 mm$^2$ sensitive to alpha particles. The sensor was connected to an electronic chain consisting of a preamplifier, amplifier, discriminator, timer and an Arduino UNO microcontroller connected to a personal computer so that information of the number of counts could be extracted and analysed.

The sensor's sensitivity to alpha particles was established under laboratory conditions using $^{241}$Am and $^{226}$Ra radioactive sources and radon gas exhaled by rocks containing uranium ore. A one-month data collection in a radon-rich atmosphere was then obtained. The correlation between the prototype monitor counts and the counts obtained with a Geiger-Müller detector, used for control, is very good. No attempt was made to calibrate the device in a standard radon source in the present work. This could be a step to be taken in a future development of the device. Nevertheless, the developed prototype has value as a didactic device, since it is an open system and can be easily used by students to learn about radon detection.

**Acknowledgments**

We are grateful to Ashley Rose Peralta for the review of the English text.